\documentstyle[aps,prd,epsf]{revtex}
\newcommand{\beq}{\begin{equation}}
\newcommand{\eeq}{\end{equation}}
\newcommand{\bd}{\begin{displaymath}}
\newcommand{\ed}{\end{displaymath}}
\newcommand{\bear}{\begin{eqnarray}}
\newcommand{\ear}{\end{eqnarray}}
\newcommand{\earn}{\nonumber \end{eqnarray}}
\newcommand{\dst}{\displaystyle}
\newcommand{\nn}{\nonumber \\}
\newcommand{\Ref}[1]{(\ref{#1})}
\newcommand{\f}{\frac}
\newcommand{\bml}{\begin{mathletters}}\newcommand{\eml}{\end{mathletters}}
\newcommand{\cZ}{{\cal Z}}

\begin{document}
%--------------------------------------------------------------------
%\twocolumn[

\title{Ground state energy in a wormhole space-time.}

\author{Nail R. Khusnutdinov\footnotemark\thanks{e-mail: nail@dtp.ksu.ras.ru},
Sergey V. Sushkov\footnotemark\thanks{e-mail: sushkov@kspu.kcn.ru}}

\address{${}^{*}$Department of Physics,
${}^{\ddag}$Department of Mathematics,\\
Kazan State Pedagogical University, Mezhlauk 1, Kazan 420021, Russia}

\maketitle

\begin{abstract}
The ground state energy of the massive scalar field with
non-conformal coupling $\xi$ on the short-throat flat-space
wormhole background is calculated by using zeta renormalization
approach. We discuss the renormalization and relevant heat kernel
coefficients in detail. We show that the stable configuration of
wormholes can exist for $\xi > 0.123$. In particular case of
massive conformal scalar field with $\xi=1/6$, the radius of
throat of stable wormhole $a\approx 0.16/m$. The self-consistent
wormhole has radius of throat $a\approx 0.0141\/ l_p $ and mass of
scalar boson $m\approx 11.35\/ m_p$ ($l_p$ and $m_p$ are the
Planck length and mass, respectively).

\vskip12pt\noindent
{\normalsize PACS number(s): 04.62.+v, 04.70.Dy, 04.20.Gz}
\end{abstract}

%--------------------------------------------------------------------
\section{Introduction}
%--------------------------------------------------------------------
Wormholes are topological handles in space-time linking widely separated regions
of a single universe, or ``bridges'' joining two different space-times. Interest
in these configurations dates back at least as far as 1916 \cite{Flamm} with
punctuated revivals of activity following both the classic work of Einstein and
Rosen in 1935 \cite{Eins} and the later series of works initiated by Wheeler in
1955 \cite{Wheeler}.  More recently, a fresh interest in the topic has been
rekindled by the works of Morris and Thorne \cite{MT} and Morris, Thorne, and
Yurtsever \cite{MTY}. These authors constructed
and investigated a class of objects they referred to as ``traversable
wormholes''. Their work led to a flurry of activity in wormhole
physics \cite{activ}.

The central feature of wormhole physics is the fact that
traversable wormholes are accompanied by unavoidable violations of the null
energy condition, i.e., the matter threading the wormhole's throat has to be
possessed of ``exotic'' properties. The classical matter does satisfy the usual
energy conditions, hence wormholes cannot arise as solutions of classical
relativity and matter. If they exist, they must belong to the realm of
semi-classical or perhaps quantum gravity.  In the absence of the complete theory
of quantum gravity, the semi-classical approach begins to play the most important
role for examining wormholes. However, there are not so much results concerning
quantized fields on the wormhole background.  Recently the self-consistent
wormholes in the semi-classical gravity were studied numerically in our works
\cite{ourwork}. As well some arguments in favor of the possibility of existence
of self-consistent wormhole solutions to the semi-classical Einstein equations
have been given in works of Khatsymovsky in Ref.\cite{Kha}.

Note that all the mentioned results were obtained within the framework of
various approximations, whereas no one up to now has succeeded in exact
calculations of vacuum expectation values on the wormhole background. The reason
for this state of affairs consists in considerable mathematical difficulties
which one faces with trying to quantize a physical field on the wormhole
background.
To overcome these difficulties, in this work we will consider a simple model of
the wormhole space-time: the short-throat flat-space wormhole. The model
represents two identical copies of Minkowski space with excised from each copy
spherical regions, and with boundaries of those regions are to be identified.
The space-time of this model is everywhere flat except a two-dimensional singular
spherical surface. Due to this fact it turns out to be possible to construct the
complete set of wave modes of the massive scalar field and calculate the ground
state energy.

The aim of our work is to calculate the ground state energy of the scalar field
on the short-throat flat-space wormhole background using the zeta function
regularization approach \cite{DowCri76,ZetaBook} which was developed in Refs.
\cite{Method,KhuBor99,BezBezKhu01}. In framework of this approach, the ground
state energy of scalar field $\phi$ is given by
\beq
E(s) = \f 12 \mu^{2s} \zeta_{\cal L} \left(s - \f 12\right),  \label{DefZeta}
\eeq
where
\beq
\zeta_{\cal L} (s) = \sum_{(n)} \left(\lambda_{(n)}^2 + m^2 \right)^{-s}
\eeq
is the zeta function of the corresponding Laplace operator.
To make the eigenvalues $\lambda_{(n)}^2$ discrete we assume the field $\phi$ to
be put into a large ball with Dirichlet boundary condition. The
$\lambda_{(n)}^2$ are eigenvalues of the three dimensional Laplace operator
${\cal L}$
\beq
(\triangle - \xi R) \phi_{(n)} = \lambda_{(n)}^2 \phi_{(n)},
\eeq
where $R$ is the curvature scalar.

The expression \Ref{DefZeta} is divergent in the limit $s\to 0$ which we are
interesting in. For renormalization we subtract from \Ref{DefZeta} the divergent
part of it
\beq
E^{\rm ren} = \lim_{s\to 0} \left(E(s) - E^{\rm div} (s)\right), \label{ERen}
\eeq
where
\beq
E^{\rm div} (s) = \lim_{m\to \infty} E(s).
\eeq
By virtue of the heat kernel expansion of zeta function is the asymptotic
expansion for large mass, the divergent part has the following form
\bear
E^{\rm div} (s) &=& \f 12\left(\f\mu m\right)^{2s} \f 1{(4\pi)^{3/2} \Gamma (s -
\f 12)} \\
&\times& \left\{ B_0 m^4\Gamma (s-2) + B_{1/2}m^3\Gamma (s-\f 32) + B_1 m^2
\Gamma (s-1) + B_{3/2} m \Gamma (s-\f 12) + B_2 \Gamma (s) \right\}, \nonumber
\ear
where $B_\alpha$ are the heat kernel coefficients.

In this case the renormalized ground state energy  \Ref{ERen} obeys the
normalization condition
\beq
\lim_{m\to \infty} E^{\rm ren} = 0.
\eeq

The organization of the paper is as follows. In Sec. \ref{Sec2} we describe a
space-time of wormhole in the short-throat flat-space approximation and analyze
the solution of equation of motion for massive scalar field. In Sec. \ref{Sec3}
we obtain close expression  for zeta function and ground state energy and calculate
corresponding heat kernel coefficients. We analyze also the expression for
ground state energy for different radius of throat. In Sec \ref{Sec4} we discuss
our results. The Appendixes A and B contain some technical details of
calculations.

We use units $\hbar = c = G = 1$ (except Sec \ref{Sec4}). The
signature of the space-time, the sign of the Riemann and Ricci
tensors, is the same as in the book by Hawking and Ellis
\cite{HawEllBook}.

%--------------------------------------------------------------------
\section{A traversable wormhole: the short-throat flat-space approximation}
\label{Sec2}
%--------------------------------------------------------------------
In this section we consider a simple model of a traversable wormhole. Assume
that the throat of the wormhole is very short, and that curvature in the regions
outside the mouth of the wormhole is relatively weak. An idealized model of such
a wormhole can be constructed in the following manner: Consider two copies of
Minkowski space, ${\cal M}_+$ and ${\cal M}_-$, with the spherical coordinates
$(t,r_{\pm},\theta_{\pm},\varphi_{\pm})$ [Notice:  ${\cal M}_+$ and ${\cal M}_-$
have a common time coordinate $t$. One may interpret this fact as the
identification $t_+\leftrightarrow t_-$.]; excise from each copy the spherical
region $r_{\pm}<a$, where $a$ is a radius of sphere; and then identify the
boundaries of those regions:
$(t,a,\theta_+,\varphi_+)\leftrightarrow(t,a,\theta_-,\varphi_-)$.
The Riemann tensor for this model is identically zero everywhere except at the
wormhole mouths where the identification procedure takes place. Generically,
there will be an infinitesimally thin layer of exotic matter present at the
mouth of the wormhole.

Such an idealized geometry can be described by the following metric
\beq\label{metric}
ds^2=-dt^2+d\rho^2+r^2(\rho)\,(d\theta^2+\sin^2\theta\,d\varphi^2 ),
\eeq
where $\rho$ is a proper radial distance, $-\infty<\rho<\infty$, and the shape
function $r(\rho)$ is
\beq
r(\rho)=|\rho\,|+a.
\eeq
%=\left\{
%\begin{array}{rcl}
%\rho+a&\quad{\rm if}& \rho>0\\
%a&\quad{\rm if}&\rho=0\\
%-\rho+a&\quad{\rm if}&\rho<0
%\end{array}
%\right. .
%\eeq
It is easily to see that in two regions ${\cal R}_+\!:\,\rho>0$ and ${\cal R}_-
\!:\,\rho<0$ separately, one can introduce a new radial coordinate
$r_\pm=\pm\rho+a$ and rewrite the metric \Ref{metric} in the usual spherical
coordinates:
$$
ds^2=-dt^2+dr_\pm^2+r_\pm^2(d\theta^2+\sin^2\theta\,d\varphi^2 ),
$$
This form of the metric explicitly indicates that the regions ${\cal
R}_+\!:\,\rho>0$ and ${\cal R}_-\!:\,\rho<0$ are flat. However, note that such
the change of coordinates $r=|\rho\,|+a$ is not global, because it is ill
defined at the throat $\rho=0$. Hence, as was expected, the space-time is curved
at the wormhole throat. To illustrate this we calculate the scalar curvature
$R(\rho)$ in the metric \Ref{metric}:
\beq
R(\rho)=-8a^{-1}\delta(\rho)\label{ScaCur}.
\eeq

Let us now consider a scalar field $\phi$ in the space-time with the metric
\Ref{metric}. The equation of motion of  the scalar field is
\beq\label{eqmotion}
(\Box-m^2-\xi R)\phi=0,
\eeq
where $m$ is a mass of the scalar field, and $\xi$ is an arbitrary coupling with
the scalar curvature $R$. In the metric \Ref{metric}, a general solution to the
equation \Ref{eqmotion} can be found in the following form:
\beq
\phi(t,\rho,\theta,\varphi)=e^{-i\omega t}u(\rho)Y_{ln}(\theta,\varphi),
\eeq
where $Y_{ln}(\theta,\varphi)$ are spherical functions, $l=0,1,2,\dots$,
$n=0,\pm1,\pm2,\dots,\pm l$, and a function $u(\rho)$ obeys the radial equation
\beq\label{radialeq}
u''+2\frac{r'}{r}u'+\left(\omega^2-\frac{l(l+1)}{r^2}-m^2-\xi R\right)u=0,
\eeq
where a prime denotes the derivative ${d}/{d\rho}$. In the flat regions ${\cal
R}_\pm$, where $r(\rho)=\pm\rho+a$, $r'(\rho)=\pm 1$, and $R(\rho)=0$,
Eq.\Ref{radialeq} reads
\beq
u''+\frac{2}{\rho\pm a}u'+\left(\omega^2-m^2-\frac{l(l+1)}{(\rho\pm
a)^2}\right)u=0.
\eeq
A general solution of this equation can be written as
\beq\label{sol}
u^{\pm}_l [\lambda (\rho\pm a)]=A^{\pm}_l h^{(1)}_l[\lambda (\rho\pm a)]+
            B^{\pm}_l h^{(2)}_l [\lambda (\rho\pm a)],
\eeq
where
$$\lambda=\sqrt{\omega^2-m^2},\quad |\omega|>m,$$
$h^{(i)}_l[z]$ are spherical Hankel functions, and $A^{\pm}_l$, $B^{\pm}_l$ are
arbitrary constants.

The solutions $u^{\pm}_l [\lambda (\rho\pm a)]$ have been obtained in the flat
regions ${\cal R}_\pm$ separately. To find a solution in the whole space-time we
must impose matching conditions for $u^{\pm}_l [\lambda (\rho\pm a)]$ at the
throat $\rho=0$. The first condition demands that the solution has to be
continuous at $\rho=0$. This gives
$$
u^{-}_l [-\lambda a]=u^{+}_l [\lambda a],
$$
or
\beq\label{cond1}
A^{-}_l h^{(1)}_l [-\lambda a]+B^{-}_l h^{(2)}_l [-\lambda a]-
A^{+}_l h^{(1)}_l [\lambda a]-B^{+}_l h^{(2)}_l [\lambda a]=0.
\eeq
To obtain the second condition we integrate Eq.\Ref{radialeq} within the
interval $(-\epsilon,\epsilon)$ and then go to the limit $\epsilon\to0$. Taking
into account the following relations\bear
&r(\rho)=|\rho\,|+a, \quad r'(\rho)=\mathop{\rm sign}\rho, \quad
r''(\rho)=2\delta(\rho),&
\nonumber\\[6pt]
&\dst\lim_{\epsilon\to0}\int_{-
\epsilon}^{\epsilon}f(\rho)\delta(\rho)d\rho=f(0),&
\nonumber
\ear
we find
\beq\label{cond}
\left.\frac{du_l^{-}[x]}{dx}\right|_{x=-\lambda a}=
\left.\frac{du_l^{+}[x]}{dx}\right|_{x=\lambda a}+\frac{8\xi}{\lambda
a}u_l^{+}[\lambda a].
\eeq
Substituting Eq. \Ref{sol} into \Ref{cond} gives
\beq\label{cond2}
A_l^{-}{h_l^{(1)}}'[-\lambda a] + B_l^{-}{h_l^{(2)}}'[-\lambda a]-
\left({h_l^{(1)}}'[\lambda a] + \frac{8\xi}{\lambda a}h_l^{(1)}[ \lambda
a]\right)A_l^{+} - \left({h_l^{(2)}}'[\lambda a] +  \frac{8\xi}{\lambda
a}h_l^{(2)} [\lambda a]\right)B_l^{+}=0,
\eeq
where ${h_l^{(i)}}' [\pm \lambda a]=(dh_l^{(i)} [x]/dx)_{x=\pm \lambda a}$.

In addition to two matching conditions \Ref{cond1} and \Ref{cond2} we must
demand  regular behavior of the scalar field at the infinity. For this aim, we
will consider a ``box approximation'', i.e., we will assume, in an intermediate
stage of calculations, that the wormhole space-time has an finite radius $R$, so
that $|\rho| \le R$, and we will go, in the end, to the limit $R\to\infty$. In
the framework of the box approximation, we demand that the scalar field becomes
to be equal zero at the space-time bounds $\rho=\pm R$. Taking into account Eq.
\Ref{sol} gives
$$
u_l^{-} [-\lambda (R+a)]=0, \quad u_l^{+} [\lambda (R+a)] = 0,
$$
or
\beq\label{cond3}
A_l^{-}h_l^{(1)} [-\lambda (R+a)]+B_l^{-}h_l^{(2)} [-\lambda (R+a)] = 0,
\eeq
\beq\label{cond4}
A_l^{+}h_l^{(1)} [\lambda (R+a)] + B_l^{+}h_l^{(2)} [\lambda (R+a)]=0.
\eeq

The four conditions \Ref{cond1}, \Ref{cond2}, \Ref{cond3} and \Ref{cond4}
obtained represent a homogeneous system of linear algebraic equations for four
coefficients $A_l^{\pm}$, $B_l^{\pm}$. As is known, such a system has a
nontrivial solution if and only if the matrix of coefficients is degenerate.
Hence we get\beq\label{det}
\left|
\begin{array}{cccc}
h_l^{(1)} [-\lambda a] & h_l^{(2)} [-\lambda a] & -h_l^{(1)} [\lambda a] & -
h_l^{(2)}
[\lambda a] \\
{h_l^{(1)}}' [-\lambda a] & {h_l^{(2)}}' [-\lambda a] & \dst
-{h_l^{(1)}}' [\lambda a] - \frac{8\xi}{\lambda a}h_l^{(1)} [\lambda a] & \dst
-{h_l^{(2)}}' [\lambda a]-\frac{8\xi}{\lambda a}h_l^{(2)} [\lambda a] \\
h_l^{(1)} [-\lambda (R+a)] & h_l^{(2)} [-\lambda (R+a)] & 0 & 0 \\
0 & 0 & h_l^{(1)} [\lambda (R+a)] & h_l^{(2)} [\lambda (R+a)]
\end{array}
\right|=0.
\eeq
After some algebra one can show that the determinant in the above formula is
factorized, and so Eq. \Ref{det} can be reduced to the following two relations:
\beq\label{spectr1}
\Psi^1_l [\lambda ]=0,
\eeq
and
\beq\label{spectr2}
\Psi^2_l [\lambda]=0,
\eeq
where the functions $\Psi_1 [\lambda ]$, $\Psi_2 [\lambda ]$ are defined as
follows:
\bear
\Psi^1_l [\lambda ]&\equiv&
\f{i\lambda}2 \sqrt{a(a+R)} \left\{ h_l^{(1)}[\lambda(R+a)] h_l^{(2)}[\lambda a]
- h_l^{(2)}[ \lambda (R+a)] h_l^{(1)}[ \lambda a]\right\}, \label{CondReal1}\\
\Psi^2_l [\lambda ]&\equiv& \f{i\lambda^2 a}8 \sqrt{a(a+R)} \left\{
h_l^{(1)}[ \lambda (R+a)]\left(\frac{4\xi}{\lambda a} h_l^{(2)}[ \lambda
a]+{h_l^{(2)}}'[\lambda a] \right) \right. \nonumber \\
&-& \left. h_l^{(2)}[ \lambda (R+a)] \left(\frac{4\xi}{\lambda a} h_l^{(1)}[
\lambda a]+{h_l^{(1)}}'[\lambda a]  \right) \right\}\label{CondReal2}.
\ear
We introduced additional factors in order to simplify formulas that follow.
These factors do  not change the relations  \Ref{spectr1}, \Ref{spectr2}.
A significance of Eqs. \Ref{spectr1} and \Ref{spectr2} is that they determine a
set of possible values of the wave number $\lambda$, i.e., a spectrum for scalar
field modes. Resolving Eq. \Ref{spectr1} and Eq. \Ref{spectr2} we can obtain two
families, respectively:
\bml \bear
&&\lambda_{lp_1}^{(1)}(a,R,\xi), \quad p_1=1,2,3,\dots, \\
&&\lambda_{lp_2}^{(2)}(a,R,\xi), \quad p_2=1,2,3,\dots.
\ear \eml

%--------------------------------------------------------------------
\section{Ground state energy and heat kernel coefficients.} \label{Sec3}
%--------------------------------------------------------------------
A ground state energy is given by
\beq
E=\frac12 \sum_{\alpha=1,2} \sum_{l=0}^\infty
     \sum_{p=1}^\infty (2 l +1 ) \sqrt{{\lambda_{lp}^{(\alpha)}}^2+m^2},
\eeq
which is, in fact, the zero point energy of massive scalar field.
This expression is divergent. In the framework of the zeta function
regularization method \cite{DowCri76,ZetaBook}, the ground state energy is
expressed in terms of the zeta function
\beq
E(s) = \f 12 \mu^{2s}\zeta_{\cal L} \left(s - \f 12\right), \label{GSE}
\eeq
where
\beq
\zeta_{\cal L} \left(s - \f 12\right) =  \sum_{\alpha=1,2} \sum_{l=0}^\infty
     \sum_{p=1}^\infty (2 l +1 ) \left( {\lambda_{lp}^{(\alpha)}}^2+m^2
\right)^{1/2 - s}
\eeq
is the zeta function associated with Laplace operator $\hat{\cal L} =
\triangle - m^2 - \xi R$. The parameter $\mu$, with dimension of mass, has been
introduced in order to have the correct dimension for the energy. For simplicity
we represent Eq. \Ref{GSE} in slightly different form
\beq
E(s) = \f 12 \left(\f\mu m\right)^{2s}\zeta \left(s - \f 12\right), \label{GSE1}
\eeq
where we introduced the function  with dimension energy
\beq
\zeta \left(s - \f 12\right) = m^{2s} \zeta_{\cal L} \left(s - \f 12\right)
\eeq
which we shall call the zeta function, too.

The solutions $\lambda_{lp}^{(\alpha)}(a,R,\xi)$ of Eqs. \Ref{spectr1},
\Ref{spectr2} cannot be found in closed form. For this reason we use the method
suggested in Ref. \cite{Method}, which allows us to express the zeta function in
terms of the eigenfunctions. The sum over $p$ may be converted into a contour
integral in a complex $\lambda$-plane using the principal of argument, namely
\beq
\zeta \left(s - \f 12\right) = \f{m^{2s}}{2\pi i} \sum_{\alpha=1,2}
\sum_{l=0}^\infty (2 l
+ 1 ) \int_\gamma d\lambda \left( \lambda^2 + m^2 \right)^{1/2 - s}
\f\partial{\partial \lambda} \ln \Psi^\alpha_l [\lambda ],
\eeq
where the contour $\gamma$ runs counterclockwise and must enclose all solutions
of Eqs. \Ref{spectr1}, \Ref{spectr2}. Shifting the contour to the imaginary
axis, we obtain the following formula for the zeta function
\beq
\zeta \left(s - \f 12\right) = - m^{2s} \f{\cos \pi s}\pi \sum_{\alpha=1,2}
\sum_{l=0}^\infty (2 l +1 ) \int_m^\infty dk \left( k^2 - m^2 \right)^{1/2 - s}
\f\partial{\partial k} \ln \Psi^\alpha_l [i k ], \label{Zeta1}
\eeq
where the functions \Ref{CondReal1}, \Ref{CondReal2} in the imaginary axis
$\lambda = ik$ read
\bml
\bear
\Psi^1_l [i k]&=&
I_\nu [k(R+a)] K_\nu [ka] - K_\nu [k(R+a)] I_\nu [ka], \\
\Psi^2_l [i k]&=& \left(\xi - \f 18\right)\Psi^1_l [ik] + \f{ka}4 \left\{
I_\nu [k(R+a)] K_\nu'[ka] - K_\nu [k(R+a)] I_\nu'[ka]\right\},
\ear
\eml
with
$$
\nu = l + \f 12.
$$
The expression \Ref{Zeta1} may be simplified in the large box limit $R \gg a$,
which we are interesting in. Let us rewrite $\Psi_l^1 [ik]$ in the following
form
\beq
\Psi^1_l [i k]=
I_\nu [k(R+a)] K_\nu [ka]\left(1 - \f{K_\nu [k(R+a)] I_\nu [ka]}{
I_\nu [k(R+a)] K_\nu [ka]}\right).
\eeq
In the large box limit, the second term in brackets obeys the inequality
\beq
\f{K_\nu [k(R+a)] I_\nu [ka]}{ I_\nu [k(R+a)] K_\nu [ka]}
< e^{-2mR}
\eeq
and brings exponentially small contribution for ground state energy.

Therefore, in the limit of large box we have
\bml
\bear
\Psi^1_l [i k]&\approx&
I_\nu [k(a + R))] K_\nu [ka], \\
\Psi^2_l [i k]&\approx& I_\nu [k(a + R)] \left\{\left(\xi - \f
18\right)K_\nu [ka] + \f{ka}4 K_\nu'[ka]\right\}.
\ear
\eml

In this moment we have to make comment on above formulas. Our approach is valid
in the case if the functions $\Psi^m_l$ in the imaginary axis do not have zeros
in the domain of integration in Eq. \Ref{Zeta1}. It gives the restriction for
$\xi$. The function $\Psi^1_l[ik]$ has no zeros in the imaginary axis, but
function $\Psi^2_l[ik]$ has simple zeros if $\xi > \f 14$. Indeed, by using
recurrent formulas for Bessel's function, let us represent the function
$\Psi^2_l[ik]$ in the following form
\beq
\Psi^2_l [i k] =  I_\nu [k(a + R)] \left\{\left(\xi - \f
18 - \f\nu 4 \right)K_\nu [ka] - \f{ka}4 K_{\nu - 1}[ka]\right\}.
\eeq
Since the Bessel's functions $K_\nu$ are positive, the expression in brackets
may change sign and therefore the function $\Psi^2_l[ik]$ may have zeros, if
\beq
\xi - \f 18 - \f\nu 4 >0.
\eeq
The lowest boundary for $\xi$ is $1/4$ for $l = 0$. More precisely, in this case
we have
\beq
\Psi^2_0 = \f 1{2k}\f 1{\sqrt{a(a+R)}} e^{kR} (1 - e^{-2k(a+R)}) \left(\xi - \f
14 - \f{ka}4\right) .
\eeq
As far as $k>m$, the function $\Psi^2_0$ has simple zero at point $k = (4\xi -
1)/a$ if
\beq
\xi > \f 14 + \f{ma}4 .
\eeq
For this reason in the paper we will consider ground state energy for $\xi <
1/4$. In opposite case we have to modify our approach.

Taking into account these formulas we may divide the zeta function, as well as
ground state energy \Ref{GSE1}, into two parts
\beq
\zeta \left(s - \f 12\right) = \zeta_R^{\rm ext} \left(s - \f 12\right) +
\zeta_a^{\rm int} \left(s - \f 12\right),
\eeq
where
\bear
\zeta_R^{\rm ext} \left(s - \f 12\right)&=& -\f{2 \beta_R^{2s}\cos \pi s}{\pi (a
+ R)} \sum_{l=0}^\infty \nu^{2-2s} \int_{\beta_R/\nu}^\infty dx \left(x^2 -
\f{\beta_R^2}{\nu^2}\right)^{1/2 - s} \f\partial{\partial x} 2\ln \left\{x^{-
\nu} I_\nu [\nu x]\right\}, \label{ZetaR}\\
\zeta_a^{\rm int} \left(s - \f 12\right)&=& -\f{2 \beta_a^{2s}\cos \pi s}{\pi a}
\sum_{l=0}^\infty \nu^{2-2s} \int_{\beta_a/\nu}^\infty d x \left(x^2 -
\f{\beta_a^2}{\nu^2}\right)^{1/2 - s} \nonumber \\
&\times &\f\partial{\partial x}\left(\ln\left\{x^\nu K_\nu [\nu x]\right\} + \ln
\left\{x^\nu \left(\delta K_\nu [\nu x] + \f{x\nu}4 K_\nu' [\nu
x]\right)\right\}\right). \label{Zetaa}
\ear
Here, $ \beta_R = m(a + R),\ \beta_a = ma,\ \delta = \xi -
\f 18$ and $\nu = l + \f 12$.

The first part of the zeta function \Ref{ZetaR} depends only on the size of box
with throat $R' = R + a$ and the asymptotic structure of the space time. It is
exactly twice the expression in the flat Minkowski space time without throat
\cite{Method} calculated for a massive scalar field inside a ball of radius $R'$
with the Dirichlet boundary condition. The factor two is very easy explained: we
consider scalar field living in the double-sided plane. The second part
\Ref{Zetaa} does not depend on a boundary and it depends only on the radius of
throat $a$ and non minimal coupling $\xi$. It contains information about the
space-time under consideration. The same division of zeta function into two
parts has been already observed for space-time of the thick cosmic string
\cite{KhuBor99} and the space-time of a point-like global monopole
\cite{BezBezKhu01}. Because the first part of zeta function \Ref{ZetaR} has
already been analyzed in great details, we proceed now to consideration the
second part \Ref{Zetaa}.

Both expressions \Ref{ZetaR} and \Ref{Zetaa} and ground state energy \Ref{GSE1}
are divergent in the limit $s\to 0$ which we are interesting in. According with
renormalization procedure, we have to subtract from regularized expression for
ground state energy \Ref{GSE1} all terms which survive in the limit $m \to
\infty$. This procedure corresponds to the subtraction five (three without
boundary) first terms of the DeWitt-Schwinger expansion
\cite{Method,KhuBor99,BezBezKhu01}.

Our goal now is to find in closed form the expansion of zeta function
\Ref{Zetaa} at the point $(-\f 12)$ as power series over $s$ (for arbitrary
mass). For this reason  we use the uniform asymptotic series over inverse index
for Bessel functions of large index and argument given in Ref. \cite{Abram}. We
subtract from and add to the integrand of Eq. \Ref{Zetaa} the uniform expansion
of it up to terms proportional to $\nu^{-3}$. After subtraction we may tend
$s\to 0$. Second part, which is the uniform expansion of integrand, gives us the
pole structure of zeta function. Going in this way (see details in Appendix
\ref{A}) we obtain the following series for zeta function at the point $(-\f
12)$
\bear
\zeta_a^{\rm int} \left(s - \f 12\right)_{s\to 0}
&=&\f 1{(4\pi)^{3/2}a\Gamma (s-\f 12)}\left\{ b_0^a\beta_a^4\Gamma (s-2)
+b_{1/2}^a\beta_a^3\Gamma (s-\f 32)+b_1^a\beta_a^2 \Gamma (s-
1)\right.\label{Zetaadimless}\\
&+&\left.b_{3/2}^a\beta_a\Gamma (s-\f 12)+b_2^a\Gamma (s)\right\} - \f 1{16\pi^2
a}\left\{b^a\ln \beta_a^2 +
\Omega^a[\beta_a]\right\}, \nonumber \\
\zeta_R^{\rm ext} \left(s - \f 12\right)_{s\to 0}
&=&\f 1{(4\pi)^{3/2}(a+R)\Gamma (s-\f 12)}\left\{ b_0^R\beta_R^4\Gamma (s-2)
+b_{1/2}^R\beta_R^3\Gamma (s-\f 32) + b_1^R \beta_R^2 \Gamma (s-
1)\right.\label{ZetaRdimless}\\
&+&\left. b_{3/2}^R \beta_R \Gamma (s-\f 12) + b_2^R\Gamma (s)\right\} - \f
1{16\pi^2(a+R)}\left\{b^R\ln \beta_R^2 +
\Omega^R[\beta_R]\right\} . \nonumber
\ear
Here
\bear
b_0^a&=& -\f{8\pi}3 ,\ b_{1/2}^a= 0 ,\ b_1^a = 32\pi \left[\xi - \f 16\right] ,\
b_{3/2}^a = 64\pi^{3/2}\left[\xi - \f 18\right]^2 ,\\
b_2^a&=& \f{8\pi}3\left[ 128 \xi^3 -64 \xi^2 + \f{56}5 \xi - \f{68}{105}\right]
,\ b_{5/2}^a= \f{16}3 \pi^{3/2}\left[96\xi^4 - 72\xi^3 + 21\xi^2 - \f{45}{16}\xi
+ \f{93}{640}\right], \nonumber \\
b^a &=& \f 12 b_0^a \beta_a^4 - b_1^a \beta_a^2 + b_2^a,
\ear
and
\bear
b_0^R&=& \f{8\pi}3 ,\ b_{1/2}^R = -4\pi^{3/2} ,\ b_1^R = \f{16}3\pi ,\ b_{3/2}^R
= - \f 13\pi^{3/2} ,\\
b_2^R&=& -\f{32}{315}\pi ,\ b_{5/2}^a= -\f 1{60} \pi^{3/2}, \nonumber \\
b^R &=& \f 12 b_0^R \beta_R^4 - b_1^R \beta_R^2 + b_2^R.
\ear
Above expressions \Ref{Zetaadimless} and \Ref{ZetaRdimless} contain all terms
which survive in the limit $s\to 0$. The details of calculation and a closed
form for $\Omega^a[\beta_a]$ are outlined in the Appendix \ref{A}. The function
$\Omega^\alpha[\beta_\alpha]$ tends to a constant for
$\beta_\alpha \to 0$ and $\Omega^\alpha[\beta_\alpha] = -b^\alpha \ln
\beta_\alpha^2 + \sqrt{\pi} b^\alpha_{5/2}/\beta_\alpha + O(1/\beta_\alpha^2)$
for $\beta_\alpha \to \infty$ ($\alpha = a, R$).

Comparing the above expression with that obtained by the Mellin
transformation taking the trace of heat kernel (in three dimensions),
\bear
\zeta \left(s - \f 12\right)_{m\to \infty}  &=& \f{m^{2s}}{\Gamma
(s - \f 12)} \int_0^\infty dt t^{s - 3/2} K[t]_{t\to 0} \nonumber \\
&=& \f 1{(4\pi)^{3/2}\Gamma (s-\f 12)}\left\{ B_0 m^4\Gamma (s-2)
+B_{1/2}m^3\Gamma (s-\f 32) + B_1 m^2 \Gamma (s-1)\right.\label{Zetadimless}\\
&+&\left. B_{3/2} m \Gamma (s-\f 12) + B_2 \Gamma (s) + \cdots  \right\},
\nonumber
\ear
we obtain the heat kernel coefficients:
\bear
B_0 &=& \f{8\pi}3 \left[(a+R)^3 - a^3\right],
\nonumber \\
B_{1/2} &=&  -4\pi^{3/2} (a+R)^2,
\nonumber \\
B_1 &=& 32\pi \left[\xi - \f 16\right]a + \f{16}3\pi (a+R),
\nonumber \\
B_{3/2} &=&
64\pi^{3/2}\left[\xi - \f 18\right]^2 - \f 13\pi^{3/2},\label{HKC} \\
B_2 &=& \f{8\pi}{3a}\left[128 \xi^3 -64 \xi^2 + \f{56}5 \xi - \f{68}{105}\right]
-\f{32}{315}\f\pi{(a+R)} ,
\nonumber \\
B_{5/2} &=& \f{16}{3} \f{\pi^{3/2}}{a^2} \left[
96\xi^4 - 72\xi^3 + 21\xi^2 - \f{45}{16}\xi + \f{93}{640}\right]  - \f1{60}
\f{\pi^{3/2}}{(a+R)^2}. \nonumber
\ear
Using above scheme we calculated also the coefficient $B_{5/2}$ which we will
need later for analysis. We should like to note the difference between Eqs.
\Ref{Zetaadimless}, \Ref{ZetaRdimless} and Eq. \Ref{Zetadimless}. The Eq.
\Ref{Zetadimless} is an asymptotic expansion of zeta function over inverse mass
$m\to \infty$ but the formulas \Ref{Zetaadimless}, \Ref{ZetaRdimless} are
correct for arbitrary mass $m$ and small $s \to 0$. In fact we extracted an
asymptotic (for $m\to \infty$) part of zeta function in the form
\Ref{Zetadimless} and saved a finite part of it. In the limit $m\to \infty$ the
finite part tends to zero and both formulas are in agreement. This is the reason
that the function $\Omega^\alpha[\beta_\alpha] = -b^\alpha \ln \beta_\alpha^2 +
\sqrt{\pi} b^\alpha_{5/2}/\beta_\alpha + O(1/\beta_\alpha^2)$ for $\beta_\alpha
\to \infty$ ($\alpha = a, R$).

As far as the space-time under consideration has singular two-dimensional
surface $\Sigma$ with codimension one, we cannot use standard formulas obtained
for smooth background and we have to utilize formulas obtained by Gilkey,
Kirsten and Vassilevich in Ref. \cite{GilKirVas01}.  The heat kernel
coefficients \Ref{HKC} coincide exactly with that obtained from general formulas
in three dimensions given in Ref. \cite{GilKirVas01}. We have to take into
account that extrinsic curvature tensor of surface $\Sigma$ is obtained as
covariant derivative of the {\em outward} unit normal vector $N_\alpha$:
\beq
K_{\alpha\beta} = \nabla_\alpha N_\beta .
\eeq
For this reason this vector has coordinates $N_\alpha = (0, \pm 1,0,0)$ on the
spheres $\rho = \pm R$, and
\beq
tr K = \f 2{R + a}
\eeq
in both cases. In Appendix \ref{A} we found general formulas for arbitrary heat
kernel coefficients and traced out them in manifest form up to $b_3$.

To obtain the ground state energy we have to subtract from our expressions
\Ref{GSE1}, \Ref{ZetaR}, \Ref{Zetaa} all terms which will survive in the limit
$m \to \infty$. Then we set $s = 0$ and radius of box $R\to \infty$. Therefore
we arrive to the following expression
\beq
E^{\rm ren} = - \f 1{32 \pi^2 a}\left\{b^a\ln \beta_a^2 +
\Omega^a[\beta_a]\right\}. \label{GeneralF}
\eeq
A similar general structure for the ground state energy in massless case has
been obtained firstly by Blau, Visser and Wipf \cite{DowCri76} using dimensional
consideration only and it was confirmed by detail calculations in Refs.
\cite{KhuBor99,BezBezKhu01}.

Using above-mentioned behavior of the $\Omega^a[\beta_a]$, the ground state
energy tends to zero for large radius of throat
\beq
E^{\rm ren} \approx - \f{b^a_{5/2}}{32\pi^{3/2} m a^2}, \ a\to \infty ,
\label{Eatoinfty}
\eeq
and it is divergent for small radius of throat:
\beq
E^{\rm ren} \approx - \f{b^a_2}{16\pi^2 a}\ln (ma), \ a\to 0 . \label{Eato0}
\eeq

\begin{figure}
\centerline{\epsfxsize=8truecm\epsfbox{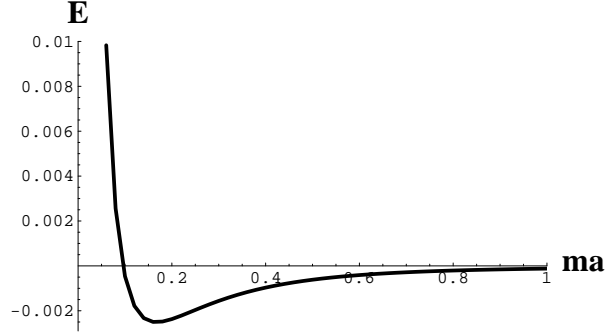}}
\caption{The ground state energy {\bf E} $= E^{ren}/m$ as a function of $ma$ for
fixed mass $m$ and $\xi = \f 16$. The energy has minimum at point $ma \approx
0.16$ with depth $E_{min}/m \approx -0.0025$}\label{f1}
\end{figure}

The numerical calculations of ground state energy $E^{ren}/m$ \Ref{GeneralF} as
a function of $\beta_a = ma$ is depicted in Fig. \ref{f1} and \ref{f2} for
$\xi = \f 16$ and $\xi = 0$, respectively. The details of numerical calculations
is analyzed in Appendix \ref{B}.

\begin{figure}
\centerline{\epsfxsize=8truecm\epsfbox{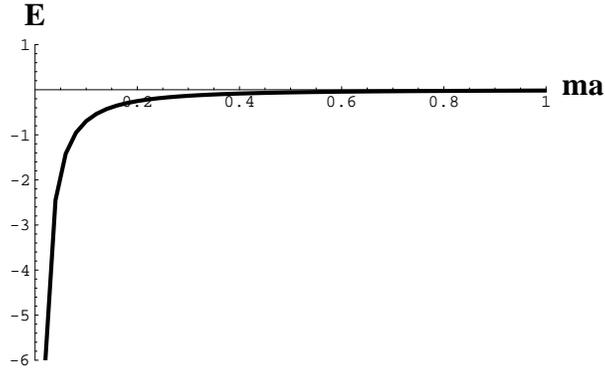}}
\caption{The ground state energy {\bf E} $= E^{ren}/m$ as a function of $ma$ for
fixed mass $m$ and $\xi = 0$. There is no minimum energy; it is always
negative.}\label{f2}
\end{figure}

%--------------------------------------------------------------------
\section{Discussion}\label{Sec4}
%--------------------------------------------------------------------
We have calculated the ground state energy of the massive scalar
field on the short-throat flat-space wormhole background (see Eq.
\Ref{GeneralF}). It can be written down in the form\footnote{In
this section we use dimensional units $G$, $c$, and $\hbar$.}
\beq\label{ground}
E^{\rm ren}=-{\hbar c\over a}f(\beta_a),
\eeq
where $\beta_a=mca/\hbar$, and $f(\beta_a)$ is a function of
$\beta_a$ which has the asymptotic
\bear
&& f(\beta_a)\approx \f{b^a_2}{16\pi^2}\ln \beta_a, \quad
\beta_a\to 0, \nn && f(\beta_a)\approx \f{b^a_{5/2}}{32\pi^{3/2}
\beta_a}, \quad \beta_a\to\infty.
\earn

To characterize the behavior of the ground state energy as a
function of $\xi$ we note that the coefficient $b_{5/2}$ is
positive for all values $\xi$ and hence the ground state energy
tends to $-0$ while as $\beta_a\to\infty$. In the limit
$\beta_a\to 0$, the behavior of the ground state energy is
determined by the sign of $b_2$ (see Eq. \Ref{Eato0}) and depends
on $\xi$. For $\xi<\xi_* \approx 0.123$, the $b_2$ is negative and
the ground state energy tends to minus infinity, otherwise it
tends to plus infinity. This difference in an asymptotical
behavior at $\beta_a\to0$ results in two qualitatively different
pictures describing the behavior of the ground state energy. In
the first case $\xi<\xi_*$, the ground state energy is
monotonically increasing from $-\infty$ to $0$ and has no extremum
(see Fig. 2); while in the second case $\xi>\xi_*$, it has a
global minimum. For example in Fig. 1 the graph of $E^{ren}/m$
versus $\beta_a$ is shown for $\xi = \f 16$. It is seen that the
ground state energy has the minimum at $\beta_a\approx 0.16$ with
depth $E_{min}/m \approx -0.0025$.

Let us now speculate about the result obtained. Suppose that the
quantum massive scalar field plays the role of the ``exotic''
matter maintaining an existence of the short-throat flat-space
wormhole in a self-consistent manner. This means that the
semiclassical Einstein equations have to be satisfied,
\beq\label{ee}
G_{\mu\nu}={8\pi G\over c^4}\langle T_{\mu\nu}\rangle^{\rm ren},
\eeq
where $G_{\mu\nu}$ is the Einstein tensor, and $\langle
T_{\mu\nu}\rangle$ is the renormalized vacuum expectation values
of the stress-energy tensor of the scalar field. The total energy
in a static space-time is given by
\beq
E=\int_V \varepsilon\, \sqrt{g^{(3)}} d^3x,
\eeq
where $\varepsilon=-\langle T_{t}^{t}\rangle^{\rm ren}= - G_t^t
c^4/8\pi G$ is energy density, and the integral is calculated over
the whole space. In the spherically symmetric metric \Ref{metric}
we obtain
\beq
E=-{c^4\over 2G} \int_{-\infty}^{\infty} G_t^t\, r^2(\rho) d\rho.
\eeq
Using the relations $G_t^t=2r''/r + (r'^2-1)/r^2$ and
$r(\rho)=|\rho\,|+a$ we can calculate
\beq\label{total}
E=-{2c^4 a\over G}.
\eeq
Note that the total energy is negative.

In the self-consistent case the total energy must coincide with
the ground state energy of the scalar field. Equating Eqs.
\Ref{ground} and \Ref{total} gives
$$
{2c^4 a\over G}={\hbar c\over a}f(\beta_a),
$$
or
\beq
a=l_P \sqrt{f(\beta_a)\over 2},
\eeq
where $l_P=\sqrt{\hbar G/c^3}$ is the Planck length. To make
further estimations we take into account that in order to be
stable a quantum system should be in the state with minimum of
ground state energy. This requirement can be fulfilled in case
$\xi>\xi_*$. In particular, for $\xi=1/6$ the minimum
$E_{min}/mc^2 \approx -0.0025$ is achieved at
$\beta_a=mca/\hbar\approx 0.16$. This gives $f(\beta_a)\approx
4\times10^{-4}$, so that
\beq
a\approx 0.0141\, l_P,
\eeq
and
\beq
m\approx 11.35\, m_P,
\eeq
where $m_P=(\hbar c/G)^{1/2}$ is the Planck mass.

Thus, our estimations have revealed that the self-consistent
semiclassical wormhole, if exists, should possess the throat of
sub-Planckian radius, and the quantum scalar field maintaining the
wormhole's existence should have super-Planckian mass. Of course,
it should be noted that our consideration has been restricted by a
toy model of the short-throat flat-space wormhole, and so one may
expect that in more realistic models results would slightly be
changed.

Let us emphasize that the result obtained in this work for the
wormhole configuration can be generalized. Really, the behavior of
ground state energy for small \Ref{Eato0} and large
\Ref{Eatoinfty} values of the throat's radius $a$ depends only on
two dimensionless heat kernel coefficients $b_2$ and $b_{5/2}$,
respectively. Instead of the radius $a$, we could use a typical
size of system $\lambda$ (throat) and calculate the coefficients
$b_2$ and $b_{5/2}$ on the corresponding background. Now let us
consider the dimensionless ground state energy $E^{ren}/m$.
Obviously, it will only depend on the dimensionless combination
$m\lambda$, and hence the limit of large (small) mass will
correspond to limit of the large (small) size of the system. Since
for renormalization we have to subtract the fist five terms (up to
$b_2$) of expansion for large mass the ground state energy in this
limit should be proportional to the next non-vanishing term of
expansion
\beq
E^{ren} \approx \f 12 \f
1{(4\pi)^{3/2}}\f{b_{5/2}}{(m\lambda)^2}\left. \f{\Gamma (s+\f
12)}{\Gamma (s - \f 12)}\right|_{s\to 0} = - \f{b_{5/2}}{32
\pi^{3/2} (m\lambda)^2},
\eeq
which coincides with Eq. \Ref{Eatoinfty}. We would like to note
that the coefficient $b_{5/2}$ is non-zero in the limit $R\to
\infty$ for background with singular scalar curvature as it was
shown in Ref. \cite{GilKirVas01}. For smooth, non-singular
geometrical characteristics of background it is zero and we have
to take into account the next non-vanishing coefficient which is
$b_3$. In this case we have the following expression in the limit
$m\lambda \to \infty$:
\beq
E^{ren} \approx - \f{b_{3}}{32 \pi^2 (m\lambda)^3}.
\eeq

The origin of logarithmic term, as well as the behavior for small
size of system is following. The structure of poles of zeta
function does not depend on the parameters of system $m$ and
$\lambda$. The subtraction of the asymptotics for great mass
brings us the following contribution to the ground state energy
\bear
\f {(m\lambda)^{2s}-1}{2 (\lambda m)(4\pi)^{3/2}\Gamma (s-\f
12)}&\times& \left\{ b_0 (\lambda m)^4\Gamma (s-2) +
b_{1/2}(\lambda m)^3\Gamma (s-\f
32)\right. \\
&+& \left. b_1 (\lambda m)^2 \Gamma (s-1) + b_{3/2} (\lambda m)
\Gamma (s-\f 12) + b_2 \Gamma (s) \right\}_{s\to 0}, \ear where
$b_\alpha$ are dimensionless heat kernel coefficients. Taking the
limit in this formula we observe that the heat kernel coefficients
with integer indices will be survived:
\beq
-\f 1{32 \pi^2 (\lambda m)} \left(\f 12 b_0 (\lambda m)^4 - b_1
(\lambda m)^2 + b_2 \right) \ln (\lambda m)^2.
\eeq
Therefore in the limit $\lambda \to 0$ one has
\beq
E^{ren} \approx -\f{b_2 \ln (\lambda m)}{16\pi^2 (\lambda m)},
\eeq
in agreement with Eq. \Ref{Eato0}.

Therefore the necessary condition that the ground state energy
will possess a minimum is following : the coefficients $b_2$ and
the next non-vanishing coefficient ($b_{5/2}$ for singular
curvature and $b_3$ for non-singular) must be positive. If it is
so, the discussion above is valid and the self-consistent
semi-classical wormhole exist. The radius of throat of stable
wormhole and the mass of scalar boson in this case depend on the
model of wormhole and value of non-conformal coupling $\xi$.

%--------------------------------------------------------------------
\section*{Acknowledgment}
The authors would like to thank M. Bordag, J.S. Dowker, K. Kirsten and D.
Vassilevich for helpful comments on some items of paper. This work was supported
by the Russian Foundation for Basic Research grant No 99-02-17941.
%--------------------------------------------------------------------

%--------------------------------------------------------------
\appendix
\section{}\label{A}
%--------------------------------------------------------------

The uniform asymptotic expansion of the modified Bessel's functions have the
form below
\bear
K_\nu [\nu x] &=& \sqrt{\f{\pi t}{2\nu}} e^{-\nu\eta }
\sum_{k=0}^\infty \f{u_k[t]}{(-\nu)^k}, \ I_\nu [\nu x] = \sqrt{\f t{2\pi \nu}}
e^{\nu\eta } \sum_{k=0}^\infty \f{u_k[t]}{\nu^k}, \label{UniExp}\\
K'_\nu [\nu x] &=& - \sqrt{\f{\pi }{2\nu x^2 t }} e^{-\nu\eta }
\sum_{k=0}^\infty \f{v_k[t]}{(-\nu)^k}, \
I'_\nu [\nu x] = \sqrt{\f 1{2\pi\nu x^2 t }} e^{\nu\eta }
\sum_{k=0}^\infty \f{v_k[t]}{\nu^k} \nonumber
\ear
where
\bear
t& = &\f 1{\sqrt{1+x^2}},\ \eta = \sqrt{1+x^2}+\ln\f x{1+\sqrt{1+x^2}},
\nonumber  \\
u_{k+1}[t] &=& \f 12 t^2 (1 - t^2) u'_k [t] + \f 18 \int_0^t (1 - 5 t^2) u_k[t]
dt,\ u_0[t] = 1 \\
v_{k+1}[t] &=& u_{k+1}[t] + t (t^2 - 1) \left\{\f 12 u_k[t] + t u'_k[t]
\right\},\ v_0[t] = 1. \nonumber
\ear

Taking into account these formulas in Eq. \Ref{Zetaa} we obtain power series
over $s$ for zeta function. The uniform asymptotic expansion \Ref{UniExp} up to
$\nu^{-n}$ allows us to take into account terms up to $m^{3-n}$. Because we need
for all terms which survive in limit $m\to \infty$ we use uniform expansion up
to $n=3$.

Therefore we have the following expression for the zeta function
\bear
\zeta_a^{\rm int} \left(s - \f 12\right) &=&
-\f{2 \beta_a^{2s} \cos \pi s}{\pi a} \sum_{l=0}^\infty \nu^{2-2s}
\int_{\beta_a/\nu}^\infty d x \left(x^2 - \f{\beta_a^2}{\nu^2}\right)^{1/2 - s}
\nonumber  \\
&\times& \f\partial{\partial x}\left( \ln \left\{x^\nu K_\nu [\nu x]\right\} +
\ln \left\{x^\nu \left(\delta K_\nu [\nu x] + \f{x\nu}4 K_\nu' [\nu
x]\right)\right\} - \sum_{k=-1}^3(- \nu)^{-k}N_k \right)\label{ZetaaIde} \\
&-&\f{2 \beta_a^{2s} \cos \pi s}{\pi a} \sum_{l=0}^\infty \nu^{2-2s}
\int_{\beta_a/\nu}^\infty d x \left(x^2 -
\f{\beta_a^2}{\nu^2}\right)^{1/2 - s} \f\partial{\partial x}\sum_{k=-1}^3(-
\nu)^{-k}N_k , \nonumber
\ear
where functions $N_p$ may be found in closed form for arbitrary index $p$ using
simple program in package {\it Mathematica}. For $p \geq 0$ they are polynomial
of $3p$ degree and has the following form
\beq
N_p[t] = \sum_{k=0}^p a_{p,k} t^{p+2k}.
\eeq

The first five $N_p$ are listed below
\bear
N_0 &=& 0 ,\ N_{-1}= 2\eta , \nonumber \\
N_1&=&\left[4\delta   - \f 14\right]t +\f 1{12}t^3 ,\label{N123}\\
N_2&=& - 8\left[\delta - \f 18\right]^2 t^2 - 2\left[\delta - \f 18 \right]t^4 -
\f 18t^6 ,\nonumber \\
N_3&=&\f 13\left[64\delta^3-24\delta^2+\f 92\delta - \f{19}{64}\right]t^3
+\f 15\left[40\delta^2 - 25\delta +\f{169}{64}\right]t^5+\f
17\left[\f{49}2\delta -\f{329}{64}\right]
t^7 +\f{179}{576}t^9. \nonumber
\ear
We should like to note that the expression \Ref{ZetaaIde} is identical to
original one \Ref{Zetaa}. First term is finite in the limit $s\to 0$; all
divergences are contained into the second part.

Integrating over $x$ with help of integral
\beq
\int_{\beta/\nu}^\infty dx x \left(x^2 - \f{\beta^2}{\nu^2}\right)^{1/2 -s}
(1+x^2)^{-p/2} = \f{\Gamma (\f 32 -s) \Gamma (s + \f{p-3}2)}{2 \Gamma(\f p2)}
\left(\f\nu\beta\right)^{p-3 + 2s} \left(1 + \f{\nu^2}{\beta^2}\right)^{-s -
\f{p-3}2}
\eeq
and taking the limit $s\to 0$ in the first term we get
\beq
\zeta_a^{\rm int} \left(s - \f 12 \right)_{s\to 0} =
-\f 1{16\pi^2 a}A_f[\beta_a] + \f 1 {(4\pi)^{3/2}a \Gamma (s-\f 12)}\sum_{k=-
1}^3(-1)^kA_k[\beta_a]. \label{Zetaint}
\eeq
where
\bear
A_f[\beta_a] &=& 32\pi\sum_{l=0}^\infty \nu^2 \int_{\beta/\nu}^\infty dx
\sqrt{x^2 - \f{\beta^2}{\nu^2}}\f\partial{\partial x}\left(\ln K_\nu(\nu x) +
\ln\left[\delta K_\nu(\nu x)+ \f{x\nu}4 K'_\nu(\nu x)\right] \right.
\label{Af}\\
&+&\left.2\nu\eta (x)+\f 1\nu N_1 - \f 1{\nu^2}N_2 + \f 1{\nu^3}N_3
\right), \nonumber \\
A_{-1} &=& 4\pi \beta^2 \Gamma (s - \f 12) \sum_{l=0}^\infty \f{\cZ (0,l+s-
1)}{\Gamma (l+ s + 1/2)}, \\
A_p &=& -8\pi^{3/2}\beta^{1-p} \sum_{k=0}^p \f{a_{p,k}}{\Gamma (l + s + 1/2)}
\cZ (2k, s + k + \f{p-1}2), \label{Ap} \\
\cZ (p,s)&=&\Gamma (s)\sum_{l=0}^\infty\f{2\nu}{(1+\nu^2/\beta_a^2)^s}
\left(\f\nu{\beta_a}\right)^p.
\ear

The first four $A_p$ are listed below
\bear
A_0[\beta_a ]&=&0 ,\nonumber \\
A_1[\beta_a ]&=&-8\pi\left[\left(4\delta -\f
14\right)\cZ (0,s) +
\f 16 \cZ (2,s+1)\right] ,\nonumber \\
A_2[\beta_a ]&=&\f {4\pi^{3/2}}{\beta_a}\left[16\left(\delta - \f 18\right)^2
\cZ (0,s+\f 12) + 4 \left(\delta - \f 18\right)\cZ (2,s+\f 32) + \f 18
\cZ (4,s+\f 52)\right] , \\
A_3[\beta_a]&=&-\f{16\pi}{3\beta_a^2}\left[\left(64\delta^3-24\delta^2+\f
92\delta -\f{19}{64}\right)\cZ (0,s+1)+\f 25\left(40\delta^2-25\delta +
\f{169}{64}\right)\cZ (2,s+2)\right. \nonumber \\
&+&\left.\f 4{35}\left(\f{49}2\delta - \f{329}{64}\right)\cZ (4,s+3)
+ \f{179}{2520}\cZ (6,s+4)\right]. \nonumber
\ear

To find the heat kernel coefficients we have to take limit $m \to \infty$ in Eq.
\Ref{Zetaint}. The asymptotic expansion of $\cZ (0,q)$ over inverse power of
$\beta^2_a$ was found in Ref. \cite{BezBezKhu01}:
\beq
\cZ (0,s)= \beta^2_a \Gamma (s-1) + 2 \sum_{l=0}^\infty \f{(-1)^l}{l!} \Gamma
(l+s) \beta^{-2l}_a \zeta_H (-1 - 2l,\f 12),
\eeq
where $\zeta_H(s,a)$ is the Hurwitz zeta function
\beq
\zeta_H(s,a) = \sum_{l = 0}^\infty  (l + a)^{-s},\ s> 1. \label{HurZet}
\eeq

Other functions $\cZ (2k, s + k + \f{p-1}2)$ in \Ref{Ap} are expressed in terms
of $\cZ (0,q)$ by relation
\beq
\cZ (2k, s + k + \f{p-1}2) = \sum_{n=0}^k \f{k!}{n!(k-n)!} \f{\Gamma (k + \f{p-
1}2 + s)}{ \Gamma (n + \f{p-1}2 + s)} \cZ (0, n + \f{p-1}2 + s).
\eeq

Taking into account above formulas we obtain the following formulas for heat
kernel coefficients
\bear
b_n &=& - \f 1{\Gamma (s-2+n)} \sum_{p=0}^n \alpha_{n-p-1} (2p-1), \\
b_{n+1/2} &=&  \f 1{\Gamma (s-\f 32+n)} \sum_{p=0}^n \alpha_{n-p-1} (2p),
\ear
where $(l,p \geq 0)$
\bear
\alpha_{-1}(-1) &=& \f{8\pi}3 \Gamma (s-2),\nonumber \\
\alpha_l(-1) &=& 16\pi (-1)^l \f{\zeta_H (-1 -2l,\f 12)}{l! (1-2l)} \Gamma (s-
1+l),\nonumber \\
\alpha_l (p) &=& -8\pi^{3/2} \sum_{k=0}^p \f{a_{p,k}}{\Gamma (k + \f p2)}
z_l(p,k), \\
z_{-1} (p,k) &=& \sum_{n=0}^k (-1)^n \f{k!}{n! (k-n)!} \f{\Gamma (k+\f{p-1}2 +
s)}{ n+\f{p-3}2 + s },\nonumber \\
z_l (p,k) &=& 2\f{(-1)^l}{l!}\zeta_H (-1-2l,\f 12)\sum_{n=0}^k (-1)^n \f{k!}{n!
(k-n)!} \f{\Gamma (k+\f{p-1}2 + s)}{\Gamma (n+\f{p-1}2 + s)} \Gamma (l+n+\f{p-
1}2 + s). \nonumber
\ear

Using these formulas, one obtains the following expressions for heat kernel
coefficients
\bear
b_0^a&=& -\f{8\pi}3 ,\ b_{1/2}^a= 0 ,\ b_1^a = 32\pi \left[\xi - \f 16\right] ,\
b_{3/2}^a = 64\pi^{3/2}\left[\xi - \f 18\right]^2 , \nonumber \\
b_2^a&=& \f{8\pi}3\left[ 128 \xi^3 -64 \xi^2 + \f{56}5 \xi - \f{68}{105}\right]
,\ b_{5/2}^a= \f{16}3 \pi^{3/2}\left[96\xi^4 - 72\xi^3 + 21\xi^2 - \f{45}{16}\xi
+ \f{93}{640}\right],  \\
b_3^a &=& \f{8\pi}{3} \left[ \f{4096}5 \xi^5 - \f{4096}5 \xi^4 + \f{35584}{105}
\xi^3 - \f{1088}{15} \xi^2 + \f{848}{105}\xi- \f{144}{385}\right]. \nonumber
\ear
The coefficient $b_{k/2}$ is polynomial of $(k-1)$-th order over $\xi$. The
coefficient $b_2$ changes its sign at point $\xi_* \approx 0.123$ and $b_{5/2}$
is positive for arbitrary $\xi$.

Our problem now is to take limit $s\to 0$ in the second part of Eq.
\Ref{ZetaaIde}. Because the function $\cZ (p,s)$ with $p = 2,\ 4,\ \cdots\ $ may
be expressed in terms the $\cZ (0,s)$ only, let us analyze it in details. Let us
suppose for a moment that $\beta_a < 1$ and represent $\cZ (0,s)$ as power
series over $\beta_a$:
\beq
\cZ (0,s) = 2\beta_a^{2s}\sum_{n=0}^\infty\f{(-1)^n}{n!}\Gamma (n+s)\beta_a^{2n}
\zeta_H\left(2n+2s-1,\f 12\right). \label{Z0}
\eeq
The gamma function $\Gamma (s)$ has simple poles in points $s = 0,\ -1,\ -2,\
\cdots$ and the Hurwitz zeta function $\zeta_H (s,p)$ has simple pole only in
one point $s=1$. They have the following expansion near their poles
\beq
\Gamma (s-n)_{s\to 0} = \f{(-1)^n}{n!}\left(\f 1s + \Psi [n+1]\right) + O(s),\
\zeta_H(s+1,p)_{s\to 0}= \f 1s - \Psi [p] + O(s),
\eeq
where $\Psi[x]$ is the digamma function.

All divergences of the function $\cZ (0,s)$ \Ref{Z0} in the limit $s\to 0$ are
contained in the first two terms with $n=0,\ 1$. The rest of series is finite
and we set $s=0$ in it. Therefore we obtain the following expression
\bear
\cZ (0,s)_{s\to 0}&=&2\beta_a^{2s}\left\{\f 12\beta_a^2 \Gamma (s-1) + \f
1{24}\Gamma (s) + \f 12\beta_a^2 \left[1 - 2\gamma - 4\ln 2\right] - \f
1{12}\left[12\zeta'_R(-1) + \ln 2 \right] \right\}\\
&+& 2\sum_{n=2}^\infty\f{(-1)^n}n  \beta_a^{2n}\zeta_H\left(2n-1,\f
12\right) + O(s),  \nonumber
\ear
where $\zeta_R (s)$ is the Riemann zeta function and $\gamma$ is the Euler
constant.

The series in above formula may be analytically continued for arbitrary value of
$\beta_a$. First of all using series representation of Hurwitz zeta function
\Ref{HurZet} we represent this series in the following form
\beq
j_2(\beta_a)=\sum_{n=2}^\infty\f{(-1)^n}{n} \beta_a^{2n}\zeta_H \left(2n-
1,\f 12\right) = \sum_{l=0}^\infty\nu\left[- \ln \left( 1 + \f{\beta_a^2}{
\nu^2}\right) +  \f{\beta_a^2}{\nu^2}\right]\ . \label{j2ln}
\eeq
Then, taking into account the integral representation for logarithm
\bd
\ln x = \int_0^x \f{dt}{1+t},
\ed
and the close expression for series below
\beq
j_0(x^2) = \sum_{l=0}^\infty\f 1{\nu (\nu^2 +x^2)} = \f 1{2x^2}\left\{
\Psi \left(\f 12 -ix\right)+\Psi \left(\f 12 +ix\right) -
2\Psi \left(\f 12 \right)\right\}\ , \label{j0}\eeq
one has
\beq
j_2(\beta ) = \beta^2\int_0^1 dxx\left\{
\Psi \left(\f 12 -ix\beta\right)+\Psi \left(\f 12 +ix\beta\right) -
2\Psi \left(\f 12 \right)\right\}\ .
\eeq
The function in rhs is analytical in whole complex plane and therefore it gives
analytical continuation the series  $ j_2(\beta_a)$ for arbitrary value of
$\beta_a$. This representation of $j_2(\beta )$ is preferable for numerical
calculations.

Using the same approach for other $\cZ (p,q)$ we arrive at the following
formulas for $A_k[\beta_a]$
\bear
A_{-1}[\beta_a]&=&-\f{8\pi}3\beta_a^{2s}\left\{\f 7{160}\Gamma (s) - \f 14
\beta_a^2 \Gamma (s-1) - \beta_a^4\Gamma (s-2)\right\} + \omega_{-1}(\beta_a),
\nonumber \\
A_1[\beta_a]&=&-\f {8\pi}3 \beta_a^{2s}\left\{\left(\delta - \f 1{16}\right)
\Gamma (s) + 12\left(\delta - \f 1{48}\right)\beta_a^2\Gamma (s-1)\right\} +
\omega_1(\beta_a), \nonumber \\
A_2[\beta_a]&=& 4\pi^{3/2}\beta_a^{2s}\left\{16\delta^2 \beta_a \Gamma (s - \f
12) + \f 4{3\beta_a} \left(\delta - \f 18\right)^2 \Gamma (s + \f 12)\right\} +
\omega_2(\beta_a), \\
A_3[\beta_a]&=&-\f{8\pi}3 \beta_a^{2s} \left\{128\delta^3 - 16\delta^2 + \f
15\delta  + \f{71}{3360} \right\}\Gamma (s) + \omega_3(\beta_a), \nonumber
\ear
where
\bear
\omega_{-1}(\beta_a)& = &8\pi\left\{\left[-\f 7{160}-\f 72\zeta'(-3)+
\f 1{240}\ln 2\right] +\beta_a^2\left[2\zeta'(-1)+\f 16\ln 2+\f 14\right]
\right. \nonumber \\
&+&\left.\beta_a^4\left[\f 13\gamma - \f{13}{36}+\f 23\ln 2\right]+
j_3(\beta_a)\right\}, \nonumber \\
\omega_1(\beta_a)&=&-16\pi\left\{\left[\f 14\zeta'(-1) + \f 1{48}\ln 2 +
\f 1{144}\right] + \delta\left[-4\zeta'(-1)-\f 13\ln 2\right]\right. \nonumber
\\
&+&\left.\beta_a^2\left[\left(\f 1{12}\gamma -
\f 18 + \f 16\ln 2\right)+\delta\left(-4\gamma - 8\ln 2 + 2\right)\right]
+ 4\left(\delta - \f 1{16}\right)j_2(\beta_a) + \f 16 \beta_a^4j_0
(\beta_a^2)\right\}, \nonumber \\
\omega_2(\beta_a)& = &-8\pi^2\left\{-16\delta^2\beta_a -16\left[ \delta -
\f 18\right]^2 j_{\f 12,0}(\beta_a) - 2\left[\delta - \f 18\right]j_{\f
32,0}(\beta_a) - \f 3{32}j_{\f 52,0}(\beta_a)\right\}, \\
\omega_3(\beta_a)& = &-\f{32}3\pi\left\{\left[\f{71}{3360}\ln 2 +\f{757}{20160}
+\f{71}{6720}\gamma\right]+\delta\left[\f 1{10}\gamma - \f 45 + \f 15
\ln 2\right] + \delta^2\left[-8\gamma+8-16\ln 2\right]\right.\nonumber \\
&+&\delta^3 \left[128\ln 2 +64\gamma\right] + \left[64\delta^3
-24\delta^2+\f 92\delta -\f{19}{64}\right]j_{1,1}(\beta_a) + \f 25
\left[40\delta^2 - 25\delta +\f{169}{64}\right]j_{2,1}(\beta_a)\nonumber
\\&+&\left.\f 8{35} \left[\f{49}2\delta-\f{329}{64}\right]j_{3,1}(\beta_a)+
\f{179}{420}j_{4,1}(\beta_a)\right\}. \nonumber
\ear
Here we introduced the following notations
\bear
j_3(\beta )&=&\sum_{n=3}^\infty\f{(-1)^n}{n(n- \f 12)(n-1)}
\beta^{2n}\zeta_H \left(2n-3,\f 12\right), \\
j_{p,q}(\beta)&=&\sum_{n=1}^\infty\f{(-1)^n}{n!}\f{\Gamma (n+p)}{\Gamma (p)}
\beta^{2n}\zeta_H \left(2n+q,\f 12\right). \label{jpq}
\ear
The functions $ j_3(\beta )$ and $j_{p,q}(\beta)$ with integer $p$ and $q$ are
expressed in terms the function $ j_0(\beta )$ only, by relations
\bear
j_3(\beta ) &=& -2\beta^4\int_0^1 dxx(1-x)^2\left\{\Psi\left(\f 12 -i\beta
x\right) + \Psi\left(\f 12 +i\beta x\right)-2\Psi\left(\f
12\right)\right\},\nonumber  \\
j_{1,1}(\beta)&=&-xj_0(x)|_{x=\beta^2}, \nonumber \\
j_{2,1}(\beta)&=&-2xj_0(x)-x^2j'_0(x)|_{x=\beta^2}, \\
j_{3,1}(\beta)&=&-3xj_0(x)-3x^2j'_0(x)-\f 12x^3j''_0(x)|_{x=\beta^2},
\nonumber  \\
j_{4,1}(\beta)&=&-4xj_0(x)-6x^2j'_0(x)-2x^3j''_0(x)-\f 16x^4j'''_0
(x)|_{x=\beta^2}. \nonumber
\ear

For half-integer indexes $p$ and $q$ in Eq. \Ref{jpq} we use the integral
representation for Hurwitz zeta function from Ref. \cite{BatErd1}
\beq
\zeta_H (s, v) = \f 1{\Gamma (s)} \int_0^\infty \f{e^{-(v - 1) x} x^{s-1}dx}{e^x
- 1}
\eeq
and obtain the following formulas
\bear
j_{\f 12,0}(\beta) &=& - \beta \int_0^\infty \f{dx}{\sinh (x)} J_1 (2\beta
x),\nonumber \\
j_{\f 32,0}(\beta) &=& - \beta \int_0^\infty \f{dx}{\sinh (x)} \left\{ 3J_1
(2\beta x) - 2\beta x J_2 (2\beta x) \right\}, \\
j_{\f 52,0}(\beta) &=& - \beta \int_0^\infty \f{dx}{\sinh (x)} \left\{ 5J_1
(2\beta x) - \f{20}{3}\beta x J_2 (2\beta x) + \f{4}{3}\beta x J_3 (2\beta
x)\right\}. \nonumber
\ear

Substituting the formulas obtained  in Eq. \Ref{Zetaint} one has
\bear
\zeta_a^{\rm int} \left(s - \f 12\right)_{s\to 0} &=& \f
{\beta_a^{2s}}{(4\pi)^{3/2}\Gamma (s-\f 12)}\left\{ b^a_0 m^4\Gamma (s-2)
+ b^a_{1/2}m^3\Gamma (s-\f 32) + b^a_1 m^2 \Gamma (s - 1) \right.
\label{Zdimless}\\
&+&\left. b^a_{3/2} m \Gamma (s - \f 12) + b_2^a \Gamma (s) \right\} - \f
1{16\pi^2 a}\Omega^a (\beta_a),
\nonumber
\ear
where
\beq
\Omega^a(\beta_a )= A_f^a(\beta_a )+\sum_{k=-1}^3(-1)^k\omega_k(\beta_a ).
\eeq
In the limit $s\to 0$ an additional finite contribution appears due to
multiplication $s \ln \beta^2_a$ and gamma functions in brackets \Ref{Zdimless}.
Because the gamma function has simple poles in points $0,\ -1,\ -2,\ \cdots$,
the heat kernel coefficients with integer indexes will give the finite
contributions and we arrive at the following expression
\bear
\zeta_a^{\rm int} \left(s - \f 12\right)_{s\to 0}
&=&\f 1{(4\pi)^{3/2}a\Gamma (s-\f 12)}\left\{ b_0^a\beta_a^4\Gamma (s-2)
+b_{1/2}^a\beta_a^3\Gamma (s-\f 32)+ b_1^a\beta_a^2 \Gamma (s-
1)\right. \nonumber \\
&+&\left.b_{3/2}^a\beta_a\Gamma (s-\f 12)+b_2^a\Gamma (s)\right\} - \f 1{16\pi^2
a}\left\{b^a\ln \beta_a^2 + \Omega^a[\beta_a]\right\}.
\ear
Therefore we divided the zeta function in two parts: the asymptotic singular
part of zeta function in standard form and finite contribution.

Using the same approach for $\zeta^{\rm ext}_R (s - \f 12)$ one has
\bear
\zeta_R^{\rm ext} \left(s - \f 12\right)_{s\to 0}
&=&\f 1{(4\pi)^{3/2}(a+R)\Gamma (s-\f 12)}\left\{ b_0^R\beta_R^4\Gamma (s-2)
+b_{1/2}^R\beta_R^3\Gamma (s-\f 32)+ b_1^R\beta_R^2 \Gamma (s-
1)\right. \nonumber \\
&+&\left. b_{3/2}^R\beta_R\Gamma (s-\f 12)+ b_2^R\Gamma (s)\right\} - \f
1{16\pi^2 (a+R)}\left\{b^R\ln \beta_R^2 + \Omega^R[\beta_R]\right\}.
\ear

By virtue of the fact that in the limit $m\to \infty$ the above formulas must
give us the asymptotic expansion \Ref{Zetadimless}, the function $\Omega
[\beta]$ has the following behavior $( \beta_\alpha \to \infty\ ,\ (\alpha = a,
R))$
\bear
\Omega^\alpha[\beta_\alpha] &=& -b^\alpha \ln \beta_\alpha^2 -
\f{2\sqrt{\pi}}{\Gamma (s - \f 12)} \sum_{k=5}^\infty b^\alpha_{k/2}\left.
\beta_\alpha^{4-k} \Gamma (s + \f k2 -2)\right|_{s\to 0} \\ &=& -b^\alpha \ln
\beta_\alpha^2
+  \sum_{k=5}^\infty b^\alpha_{k/2} \beta_\alpha^{4-k} \Gamma (\f k2 -2)
\nonumber \\
&=& -b^\alpha \ln \beta_\alpha^2
+  \sqrt{\pi} b^\alpha_{5/2} \beta_\alpha^{-1} +  O(\beta_\alpha^{-2}).
\nonumber
\ear

%--------------------------------------------------------------
\section{}\label{B}
%--------------------------------------------------------------
The main problem for numerical calculation the ground state energy is the term
$A_f[\beta]$ given by Eq. \Ref{Af}. The series in \Ref{Af} is low convergent. To
calculate this expression let us, first of all, represent it in the following
form
\bear
\overline{A}_f &\equiv& \f{A_f}{32\pi} = \sum_{l=0}^\infty \sigma_\nu ,\ {\rm
where} \\
\sigma_\nu &=&  \nu^2 \int_{\beta/\nu}^\infty dx
\sqrt{x^2 - \f{\beta^2}{\nu^2}}\f\partial{\partial x}\left(\ln K_\nu(\nu x) +
\ln\left[\delta K_\nu(\nu x)+ \f{x\nu}4 K'_\nu(\nu x)\right] \right. \\
&+&\left.2\nu\eta (x)+\f 1\nu N_1 - \f 1{\nu^2}N_2 + \f 1{\nu^3}N_3
\right) \nonumber
\ear
and divide the series in two parts
\beq
\overline{A}_f = \sum_{l=0}^N \sigma_\nu + \sum_{l=N+1}^\infty \sigma_\nu .
\eeq

The first sum we calculated numerically. The calculations become lighter due to
fact that the Bessel's functions of second kind with half-integer indexes are
polynomial with simple exponent factor \cite{Abram}. In the second sum we use
the uniform expansion of the integrand over inverse power of index $\nu$. Since
we have already subtracted first three terms $N_1,\ N_2$ and $N_3$, the uniform
expansion of integrand will start from $\nu^{-4}$ and we obtain the following
expression
\beq
\sum_{l=N+1}^\infty \sigma_l = \sum_{l=N+1}^\infty \nu^2 \int_{\beta/\nu}^\infty
dx
\sqrt{x^2 - \f{\beta^2}{\nu^2}}\f\partial{\partial x} \sum_{p=4}^\infty (-
\nu)^{-p} N_p[t]\ , \label{tail}
\eeq
where $N_p[t]$ is the polynomial of $3p$ degree:
\beq
N_p[t] = \sum_{k=0}^p a_{p,k} t^{p + 2k}.
\eeq
The coefficients $a_{p,k}$ for $p=1,2,3$ may be singled out from Eq. \Ref{N123}.

Then one takes derivatives and integrals in Eq. \Ref{tail} and changes the sums
over $p$ and $l$. After this we arrive at the following formula
\beq
\sum_{l=N+1}^\infty \sigma_l =  \sum_{p=4}^\infty
\overline{N}_p, \label{UnAp}
\eeq
where
\bear
\overline{N}_p &=& \f{\sqrt{\pi}}2 (-1)^{3 - p} \sum_{k=0}^p a_{p,k} \f{\Gamma
\left(\f p2 - \f 12 + k\right)}{\Gamma\left( \f p2 + k\right)} h[p,p +
2k,\beta,N], \\
h[p,q,\beta,N] &=& \sum_{l=N+1}^\infty \nu^{2-p} \left(1 +
\f{\beta^2}{\nu^2}\right)^{-(q-1)/2}.
\ear
The function $h$ may be found in closed form for integer $p$ and $q$.

The above function $h$ may be estimated by
\beq
h[p,q,\beta,N] \approx \left(N+\f 32\right)^{2 - p},
\eeq
and the series \Ref{UnAp} is fast convergent for great $N$. We use $N=10$ and in
order to work with precision $10^{-10}$ it is enough to take expansion up to
$p=8$ (five terms). In fact, this procedure converts the low convergent series
to fast convergent series over $N^{2-p}$.

\begin{figure}
\centerline{\epsfxsize=8truecm\epsfbox{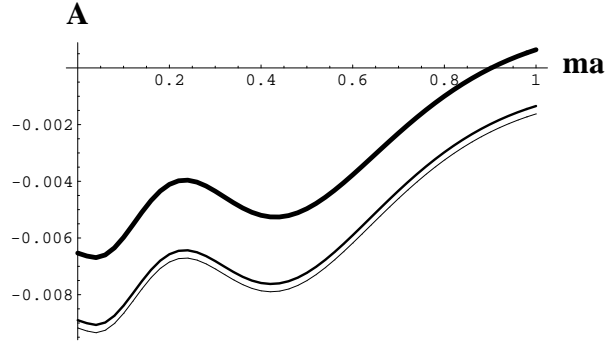}}\caption{The function {\bf A}
$= \f{A_f}{32\pi}$ as a function of $ma$ for $\xi
= \f 16$. Thick curve is zero term $(l=0)$ contribution. The curve of middle
thickness is the contribution of the first eleven terms up to $l=10$. Thin
curve reproduces the calculations with high precision (up to $p=8$ of uniform
expansion).}\label{f3}
\end{figure}

To illustrate above approach we reproduce in Fig. \ref{f3} the three steps of
calculation of $\overline{A}_f$: (i) the zero term (thick curve), (ii) the
contribution of first eleven terms up to $l=10$ (middle thickness) and (iii)
exact curve (up to $p=8$ in uniform expansion, thin curve). Ten terms $l=1\div
10$ gives us the correction of 36\% for zero term, and the series from $l=11$ to
$\infty$ gives us 4\% correction.

%--------------------------------------------------------------------


\begin{thebibliography}{99}
\bibitem{activ}
The reader can find some references on wormhole physics in the excelent
book by Visser \cite{VisserBook}.
\end{thebibliography}
\end{document}